\documentclass[12pt]{article}
\usepackage{amssymb,amsmath}
\usepackage{epsfig}
\usepackage{graphicx}

\def\shiftdown#1{#1\llap{\lower.04ex\hbox{#1}}}

\begin{document}
\begin{center}
{\Large\bf \boldmath Scattering Processes of Excited Exotic Atoms:
Close-Coupling Approach} 

\vspace*{6mm}
{V.P. Popov and V.N. Pomerantsev }\\      
{\small \it Institute of Nuclear Physics, Moscow State University,
119992 Moscow, Russia$^a$ Institution1 }      
\end{center}

\vspace*{6mm}

\begin{abstract}

The scattering processes of exotic atoms in excited states from
hydrogen such as elastic scattering, Stark transitions and Coulomb
de-excitation are studied within a close coupling approach. The
vacuum polarization and the strong interaction shifts of
$ns$-states (in case of hadronic atoms) are taken into account.
The differential and integral cross sections of the above
processes are calculated to use them as the input in cascade
calculations. The effect of closed channels on the scattering
processes is investigated.
\end{abstract}

\vspace*{6mm}
\section{Introduction}
The exotic hydrogen--like atoms are formed in highly excited states,  
when negative particles ($\mu^-, \pi^-, K^-...$) are stopped in hydrogen. 
The deexcitation of exotic atoms proceeds via many intermediate states  
until the ground state is reached or a nuclear reaction takes place.  
Despite a long history of theoretical and experimental studies 
the kinetics of this atomic cascade is not yet fully understood.  
The present experiments with the exotic 
hydrogen-like atoms addresses a number of fundamental problems using 
precision spectroscopy methods, the success of which relies crucially 
on a better knowledge of the atomic cascade.  
 The experimental data are mainly appropriate to the processes
at the last stage of the atomic cascade  (X-ray yields
and the products of the weak or strong interaction of the
exotic particle in the low angular momentum states with
hydrogen isotopes).
 So the reliable theoretical backgrounds
on the processes both in low-lying and in highly excited states are required for
the detailed and proper analysis of the experimental data.

In this paper we present the {\em ab initio}
quantum-mechanical treatment of non-reactive scattering
processes of the excited exotic hydrogen atom in collisions with the
hydrogenic atom in the ground state:
\begin{equation} \label{eq. 1}
(aX)_{n,l} + (be^-)_{1s} \to (aX)_{n',l'} + (be^-)_{1s},
\end{equation}
elastic scattering($n'\!=\!n,\,l'\!=\!l$), Stark transitions
($n'\!=\!n,\,l'\!\neq \!l$), and Coulomb deexcitation (CD) ($n'\!< \!n$). 
Here $(a,b) = (p,d,t)$ are hydrogen isotopes and $X=\mu^-,
\pi^-,K^-, \tilde{p}$; $(n,l)$ are the principal and
orbital quantum numbers of exotic atom.   While the deexcitation processes are 
obviously essential for the 
atomic cascade, the role of the collisional processes preserving 
the principal quantum number $n$ is also very important.   The Stark transitions 
affect the population of the $nl$ sublevels and 
together with the elastic scattering  they decelerate the exotic atoms 
thus influencing their energy distribution during the cascade.

 Starting from the classical paper by Leon and Bethe~\cite{LB},
 Stark transitions has been treated in the semiclassical
 straight-line-trajectory approximation (see~\cite{TH} and
 references therein). 

 The fully quantum-mechanical treatment
 of the elastic scattering and Stark transitions based on the adiabatic description
 was given for the first time in~\cite{1,2}. Recently~\cite{3},
 the processes have also been studied in a close-coupling approach
 with using dipole approximation for interaction potential and taking
 the electron screening effect into account by the model.
 The cross sections calculated in approaches~\cite{1,2,3} are in good agreement.

Concerning the CD process, the situation 
is much less defined, especially for low $n$. The first
work on the CD process was
performed by Bracci and Fiorentini~\cite{4} in frame of the
semiclassical approach with some additional approximations.
In the following numerous papers (see~\cite{AAA,AAA1} and references therein) the CD process is considered 
within the
asymptotic approaches using the adiabatic hidden crossing
theory. The CD calculations  were also performed in the classical-trajectory
Monte-Carlo (CTMC) approach\cite{7}. While the Coulomb
deexcitation cross sections obtained in CTMC approach are
in fair agreement with the semiclassical ones of Bracci and
Fiorentini~\cite{4} , the more elaborated advanced
adiabatic approach (AAA)~\cite{AAA,AAA1} gives too small CD
cross sections to explain the
experimental data~\cite{8}. The reasons of such a strong discrepancy
are not clear. One can only assume that the semiclassical
model of Bracci and Fiorentini as well as the CTMC approach are not valid for
low-lying states and at low energies. 

The processes (1) have been treated recently by authors in a
unified manner in the framework of the close-coupling (CC)
approach (see for detail~\cite{9,10,11,12}). The differential and
integral cross sections for the processes (1) have been calculated
for muonic, pionic and antiprotonic hydrogen atoms in excited
states with $n=2-14$ and in a kinetic energy range relevant for
cascade calculations. The energy shifts of the $ns$ states due to
vacuum polarization and strong interaction (for hadronic atoms)
are included in the close-coupling method. This approach allows to
obtain the self-consistent description of all the processes (1)
and is free from the additional approximations used in previous
studies. The calculated differential and integral cross sections
presented in this paper mainly refer to the CD process to
illustrate some of our new results obtained quite recently.

 All open channels corresponding to
the exotic atom states with $n\le n_0$ ($n_0$ is principal quantum
number in the entrance channel) have been included in the
CC calculations. The effect of the closed channels
with $n>n_0$ was also studied and will be discussed below.
(Throughout the whole paper the cross sections are given in atomic
units.)

\section{Close-coupling approach}

 The Hamiltonian of the system $((X^- p)_{nl}  + H_{1s})$
 (after separation of the c.m. motion) is given by
\begin{equation} \label{eq2}
H = -\frac{1}{2m}\Delta _{\mathbf{R}} +
h_{ex}(\boldsymbol{\rho}) + h_H(\mathbf{r})
+V(\mathbf{r},\boldsymbol{\rho},\mathbf{R}),
\end{equation}
where $m$ is the reduced mass of the system, $\mathbf{R}$ is the
radius vector between the c.m. of the colliding atoms,
$\boldsymbol{\rho}$ and $\mathbf{r}$ are their inner
coordinates. The interaction potential,
$V(\mathbf{r},\boldsymbol{\rho},\mathbf{R})$, is a sum of four
Coulomb pair interactions between the projectile atom and the
target atom particles. $h_{ex}(\boldsymbol{\rho})$ and
$h_H(\mathbf{r})$ are the hydrogen-like Hamiltonians of the free
exotic and hydrogen atom, whose eigenfunctions together with the
angular wave function $Y_{L\Lambda}(\hat{\bf R})$ of the
relative motion form the basis states, $|1s, n l,L:JM\rangle$,
with the conserving total angular momentum ($J M$) and parity
$\pi=(-1)^{l+L}$. In the present consideration we use the
"frozen" electron approximation. The CC approach can be extended
in a straightforward manner to include the target electron
excitations. The total wave function of the system are expanded
in terms of the basis states as follows
\begin{equation} \label{eq3}
\Psi^{J M \pi}_E(\mathbf{r}, {\boldsymbol{\rho}},
\mathbf{R}) = R^{-1} \sum_{nl L}G_{nlL}^{J \pi}(R) |1s, n
l,L:J M\rangle.
\end{equation}

The expansion (3) leads to the close-coupling second order
differential equations for the radial functions of the
relative motion, $G_{nlL}^{J \pi}(R)$,
 \begin{equation}  \label{eq4}
\left(\frac{d^2}{dR^2} + k^2_{n} -
\frac{L(L+1)}{R^2}\right)G^{J \pi }_{nlL}(R) 
=2m\sum_{n'l'L'}W^{J \pi}_{n'l'L', nlL}(R)\, G^{J
\pi}_{n'l'L'}(R).
\end{equation}
The channel wave number is defined as
$k^{2}_{n}=2m(E_{cm}+E_{n_0 l_0}-E_{n l})$, where $E_{cm}$ and
($n_0 l_0$) are the energy of the relative motion and the exotic
atom quantum numbers in the entrance channel, respectively. The
bound energy of the exotic atom, $E_{n l} = \varepsilon _{nl} +
\Delta \varepsilon _{nl},$ includes the eigenvalue of
$h_{ex}(\boldsymbol{\rho})$, $\varepsilon _{nl}$,  and the
energy shift, $ \Delta \varepsilon _{nl}=\Delta \varepsilon
^{vp} _{nl}+\Delta \varepsilon ^{str}_{nl}$, due to the vacuum
polarization and strong interaction (in case of adronic atom). Hereafter, the energy
$E_{cm}$ will be referred to $\varepsilon _{n l\neq 0}$ in the
entrance channel (we assume here that $\Delta \varepsilon _{n l\neq
0}=0$)

The matrix elements of the interaction potential,
$V(\mathbf{r},\boldsymbol{\rho},\mathbf{R})$,
\begin{equation} \label{eq5}
W^{J}_{n'l'L', nlL}(R)\!=\!\langle 1s,n'l',L'\!:\!JM|V|1s,nl,L\!:\!JM\rangle
\end{equation}
are obtained by averaging it over the electron wave function of
the $1s$-state and then applying the multipole expansion. The
integration over (${\boldsymbol{\rho}}, \hat{\bf R})$ reduces
the matrix elements (5) to the multiple finite sum.

At fixed $E_{cm}$ the coupled differential equations (4) for the
given $J$ and $\pi$ values are solved numerically by the Numerov
method with the standing-wave boundary conditions involving the
real symmetrical $K$-matrix related to $T$-matrix by the
equation $T=2iK/(I - iK)$. All open channels corresponding to the 
exotic atom states with $n\le n_0$ have been included in the
close-coupling calculations. The effect of closed channels with $n>n_0$ was also
investigated and will be discussed below.  In the next sections we present and
discuss 
the following total cross sections of the scattering processes: 
the partial cross section $\sigma^J_{nl\to n'l'}$
\begin{equation}
\sigma^J_{nl\to n'l'}  =
\frac{\pi}{k_{n}^2}\frac{2J\!+\!1}{2l\!+\!1} \sum_{ L L'
\pi}|T^{J\pi}_{nlL\to n'l'L'}|^2,\, 
\end{equation}
the total cross section of the $nl\to n'l'$ transition 
$\sigma_{nl\to n'l'}  = \sum_{J}\sigma ^J_{nl\to
n'l'}(E)$, and the  $l$-averaged cross section
$ \sigma_{nn'} =1/n^2 \sum_{l'}(2l'\!+\!1)\sigma_{nl\to n'l'}(E)$, and also 
the analogous differential cross sections.

\section{The elastic scattering and Stark transitions}

 The total cross sections
calculated in adiabatic~\cite{2} and present CC approaches are as
a whole in good agreement at energies $E_{\rm cm}>1$~eV .
  The angular distributions obtained in the present CC and
  adiabatic~\cite{1} calculations  coincide in the region of
the diffraction maximums
 but demonstrate significant differences in backward hemisphere
 (see the left side of Fig.1).
     \begin{figure}[h]
     \centerline{\includegraphics[width=0.45\textwidth,keepaspectratio]{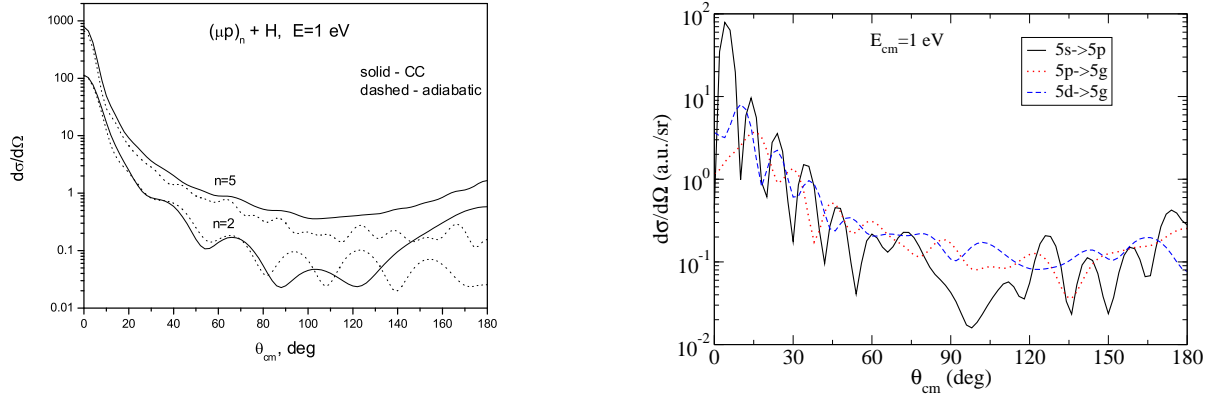}
     \hfill
     \includegraphics[width=0.45\textwidth,keepaspectratio]{fig2dd.eps}}
     \caption{The $l$-averaged differential (elastic and Stark) cross sections (left):
     adiabatic model~\cite{1}(dashed lines) and present CC (solid
          lines). The $l$-averaged differential Stark cross sections
          for $(\mu p)_{5l} + H$ collisions vs. cms
     scattering angle $\theta_{\rm cm}$ at $E_{\rm {cm}} = 1$~eV (right).}
   \end{figure}
   The typical angular distributions
  of the Stark $nl\rightarrow nl'$
transitions for $n=5$ are shown in Fig.~1 (right). It is well
known~\cite{1,3} that cross sections of these processes are
similar to the diffraction scattering (at energies more $1$~eV)
with a strong forward peak enhancing with increasing energy and a
set of maxima and minima. While the elastic cross sections always
have a strong peak at $\theta_{\rm {cm}}=0$,
 the first maximum position in the Stark cross section depends on
 the $\Delta l=|l-l'|$ value. In particular, for $\Delta l=1$ this maximum is
at finite scattering angles as it is also remarked in~\cite{3}.

 \begin{figure}[h!]
 {\includegraphics[width=0.45\textwidth,keepaspectratio]{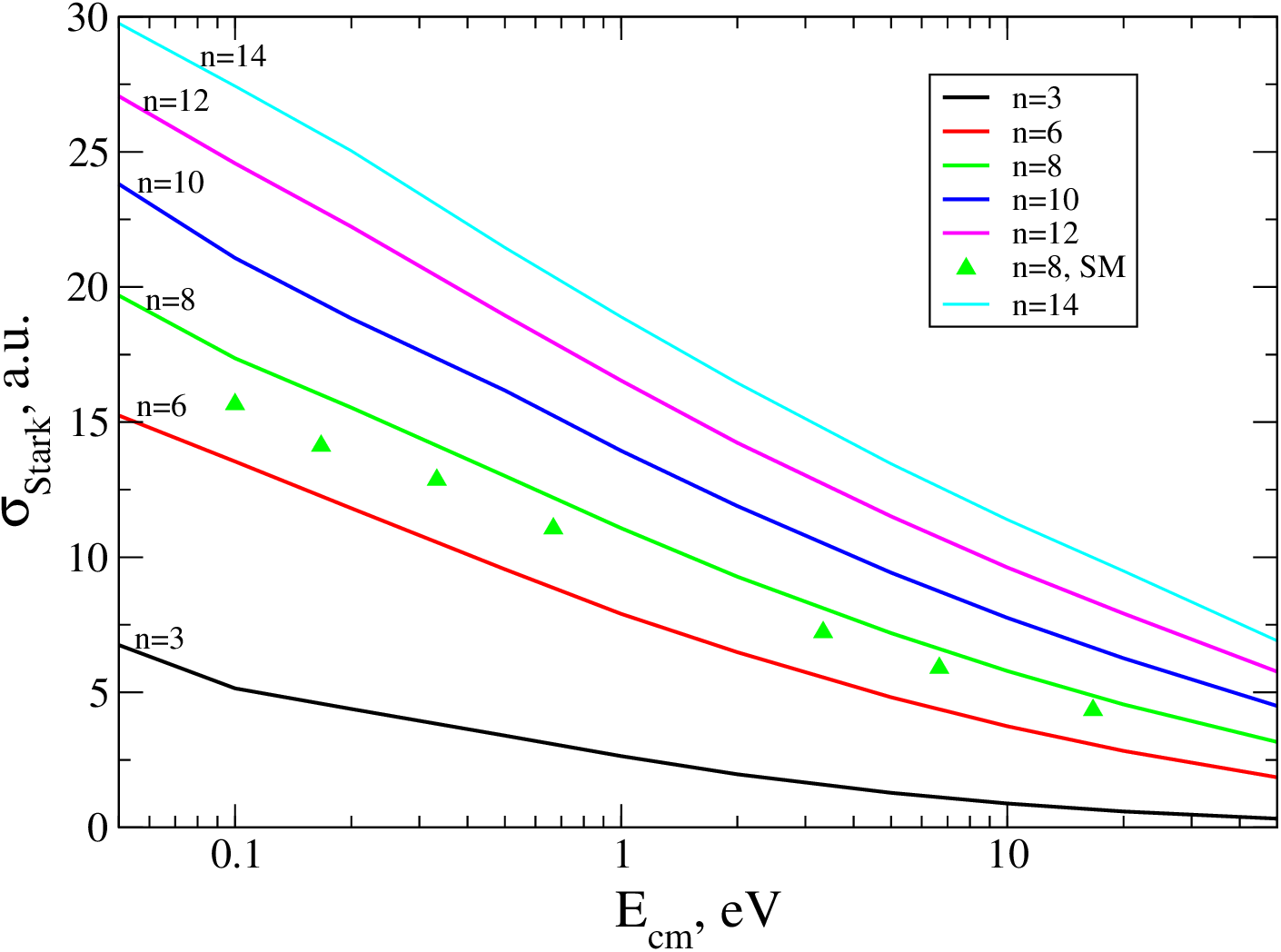}
  \hfill
  \includegraphics[width=0.45\textwidth,keepaspectratio]{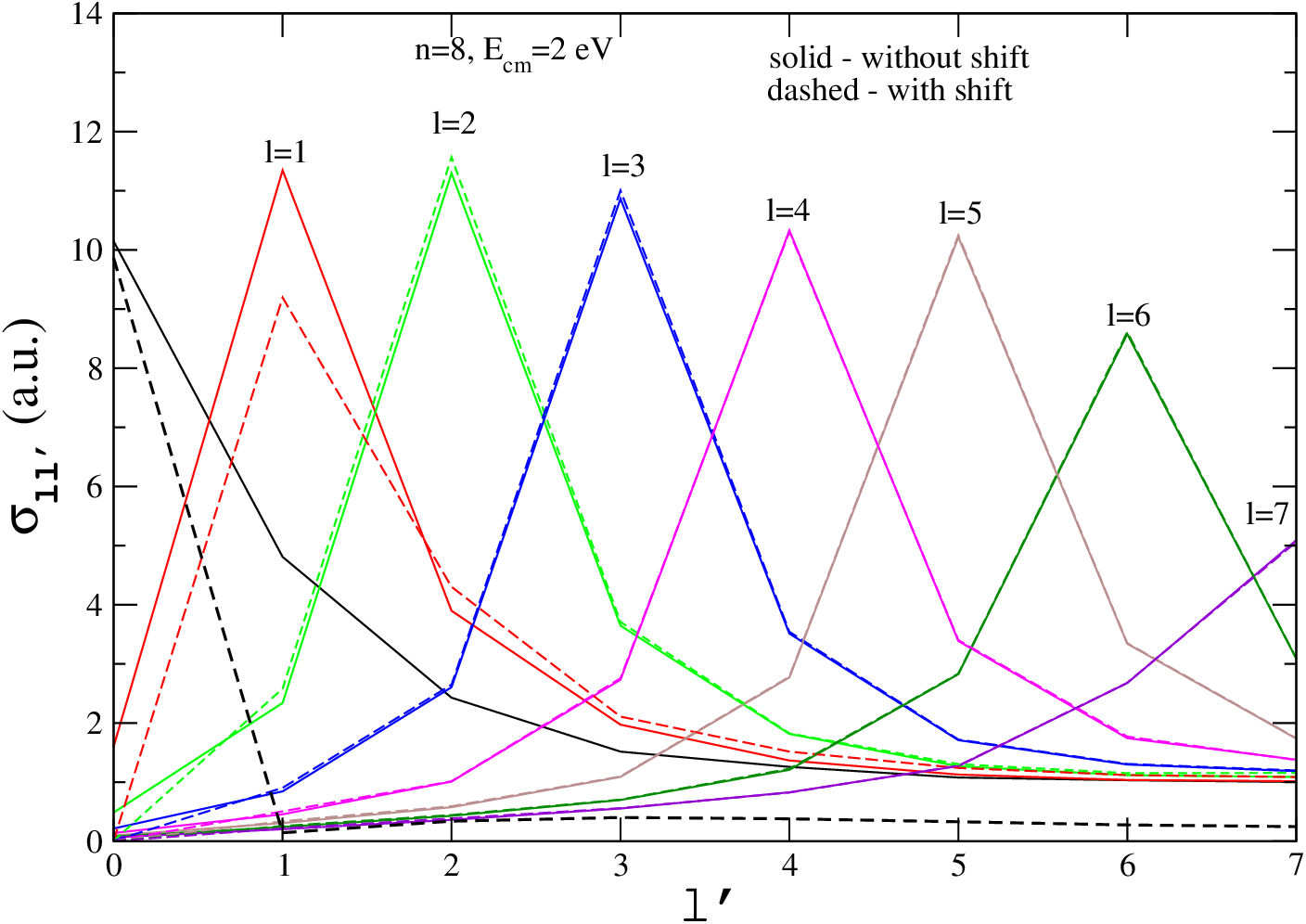}} \\
    \parbox[b]{0.45\textwidth}{\caption{The $l$-averaged Stark cross sections for
   $(p\bar{p})_n + H$ collisions(present - solid line, semiclassical
model~\cite{3} for $n=8$ - triangles)}}
  \hfill
  \parbox[b]{0.45\textwidth}{\caption{The cross sections $\sigma_{nl\to nl'}$ for
     collisions $(p\bar{p})_{n=8}+H$. The dashed and solid
     lines connect the points corresponding to the calculations both
     with and without the $ns$-state energy shifts,
     respectively.}}
     \end{figure}
 Some of our CC results for $(p\bar{p})_n +H$ collision are shown in Figs. 2,3:
 the E-dependence of $l$-averaged Stark cross sections for $n=2-14$ (Fig.2)
 and cross sections of the $nl\rightarrow nl'$ transitions for $n=8$ at $E_{\rm cm}=2$~eV
 (Fig.3). It is seen from Fig.3 that $ns$-state energy shift due to strong interaction
  leads to essential suppression of both Stark $ns\rightarrow n l>0$ and elastic $np\rightarrow np$ transitions
  (we used the value  $\epsilon_{1s} = 721$~eV for the  $1s$-state shift and $\epsilon_{ns}=\epsilon_{1s} /n^3$
 for $ns$-states) at energies compared with the shift value. The same effect is observed for all hadronic
 atoms.

\section{Coulomb de-excitation}

 The nature of the CD process is quite different from the one of the elastic or Stark processes. In contrast to elastic (Stark)
 scattering CD process is accompanied by the large energy release (tens
and hundreds eV) and occur at smaller distances, so the details of
the short-range interaction are more important for the treatment
of CD process than for elastic processes.
    \begin{figure}[h!]
     \parbox[c]{0.6\textwidth}{\includegraphics[width=0.6\textwidth,keepaspectratio]{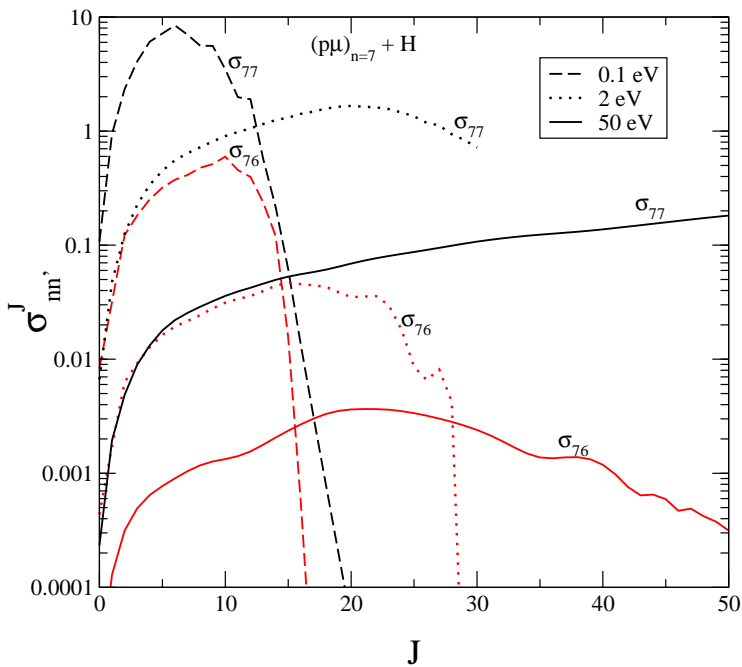}}
     \hfill
     \parbox[c]{0.35\textwidth}{\caption{The partial cross sections (in a.u.) of elastic scattering
     $\sigma^J_{7,7}$ and Coulomb de-excitation
     $\sigma^J_{7,6}$ for $(\mu p)_{n=7} + H$ collisions  versus the total angular momentum
     $J$ at the energies: 0.1~eV (dashed), 2~eV (dotted) and 50~eV
     (solid).}}
     \end{figure}
 This difference between elastic and deeply
inelastic processes is illustrated by
 Fig.~4 where the $J$ dependence of the partial-wave l-averaged
cross sections $\sigma^J_{nn'}$  for $n=7$ at three fixed energies
0.1, 2, and 50 eV is shown. It is seen that a substantial part of
the CD cross section ($\sigma^J_{76}$) comes from the partial
waves with rather a low $J$ in contrast to  the elastic (Stark)
process. 

In spite of the value of the total CD cross section
constitutes about few per cent of the total elastic cross section
it is incorrect to treat CD in the framework of the perturbation
theory. In each significant for CD partial wave the value of the
CD cross section is comparable with the elastic cross section. The
CD process is determined by the short-range behaviour of the wave
function which changes when new channels are included in
calculation. Therefore, to calculate the transition $n\to n-1$ in
a proper way it is impossible to be restricted
 with the two-level approximation ($n$ and $n-1$) and
 the states with other nearest $n$ should be involved.

 We studied the dependence of the results on the number of included
 channels and found that the inclusion of the channels
with $n<n-1$ leads to a strong suppression of the main $n\to n-1$
transitions in comparison with the two-level CC approximation and
due to this the total CD cross section is also
suppressed~\cite{9}. So in all our CD calculations we included all
the open channel with $n\le n_0$. It should be noted that all the
previous calculations of CD realized within semiclassical or
adiabatic approaches used a two-level approximation. It is obvious
that two-level approximation is  not absolutely suitable for the
treatment of transitions with $\Delta n>1$.
 In contrast to the elastic scattering and Stark
transitions where the ``dipole'' approximation ($t_m=1$) and even
more rough dipole potential (used in~\cite{3}) gives reasonable
results (at not too low energies), in case of the deeply inelastic
process such as CD the full interaction must be used as it is
clear from the present study.
\subsection{Muonic atoms}
It is commonly believed~\cite{4,AAA,AAA1} that the
 CD cross sections at low energies
behave like $1/E$. In order to reveal more explicitly the
distinction from the $1/E$ behaviour the  present $l$-averaged CD
cross sections multiplied by energy are shown in Fig.5 in
comparison with the results of the SC model~\cite{4} for
$n=3,5,7,9$ and CTMC calculations~\cite{7} for $n=9$.
\begin{figure}[h!]
     \parbox[c]{0.6\textwidth}{\includegraphics[width=0.6\textwidth,keepaspectratio]{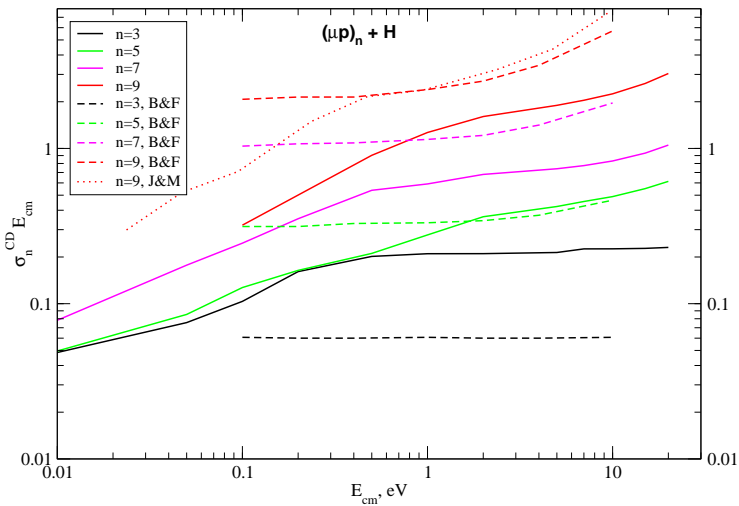}}
      \hfill
      \parbox[c]{0.35\textwidth}{\caption{The cross sections of Coulomb de-excitation
      (multiplied by $E_{\rm cm}$) for $(\mu p)_n + H$ collisions calculated in the present
      CC method (solid lines) in comparison with the SC~\cite{4} (dashed) and
       CTMC~\cite{7} results (dotted).}}
      \end{figure}
 As it is seen, the energy dependence of the CC cross
sections in the region $5> E >1$ eV, as a whole, is in a
qualitative agreement with the results~\cite{4} and \cite{7}. At
lower energies the present CC results reveal $\sim 1/\sqrt{E}$
dependence in accordance with the Wigner threshold law (the
similar behaviour is seen in the CTMC results~\cite{7} for $n=9$)
and in disagreement with $1/E$ dependence obtained in the SC
model~\cite{4} and AA approach~\cite{AAA,AAA1}. 

The distribution over
the final states $n'$ is strongly different from the SC
results~\cite{4} as illustrated in Fig.~6. The CC calculations
predict that the transitions with $\Delta n> 1$ are strongly
enhanced as compared with the results of the two-level approaches
\cite{4,AAA,AAA1}. The $\Delta n> 1$ transitions make up a substantial
fraction (16\% - 37\%) of the total CD cross section for $n\ge 4$.
     \begin{figure}[h]
    \parbox[c]{0.6\textwidth}{
     \includegraphics[width=0.6\textwidth,keepaspectratio]{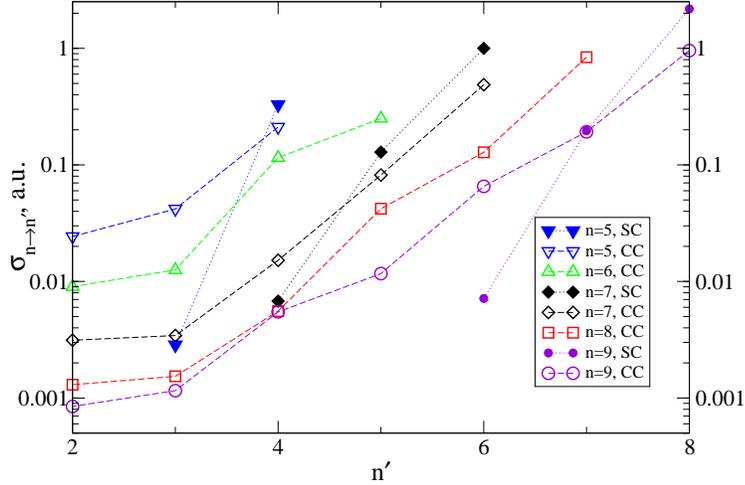}}
     \hfill
     \parbox[c]{0.35\textwidth}{\caption{Dependence of the CD cross sections on the
final principal quantum number $n'$ for the different initial $n$
in the $(\mu p)_n + H$ collisions at $E=1$ eV. The dashed and
dotted lines connect the points calculated in the present CC and
in the semiclassical (SC)~\cite{4} approaches, respectively.}}
     \end{figure}

\subsection{Differential cross sections}
The angular distributions of CD was calculated for the first time
in~\cite{11}. Earlier in the cascade calculations the angular
distributions of the CD process are presumed to be isotropic. The
calculated cross sections  for individual $nl\rightarrow n'l'$
transitions with $\Delta n=1$ and 2 at energy $E_{\rm {cm}} =
1$~eV are shown in Fig.~7.
 In Fig.~8 the $l$-averaged cross sections for the
$6\to 5$ transition at different values of the relative energy
from 0.01 up to 15~eV are presented.
     \begin{figure}[h!]
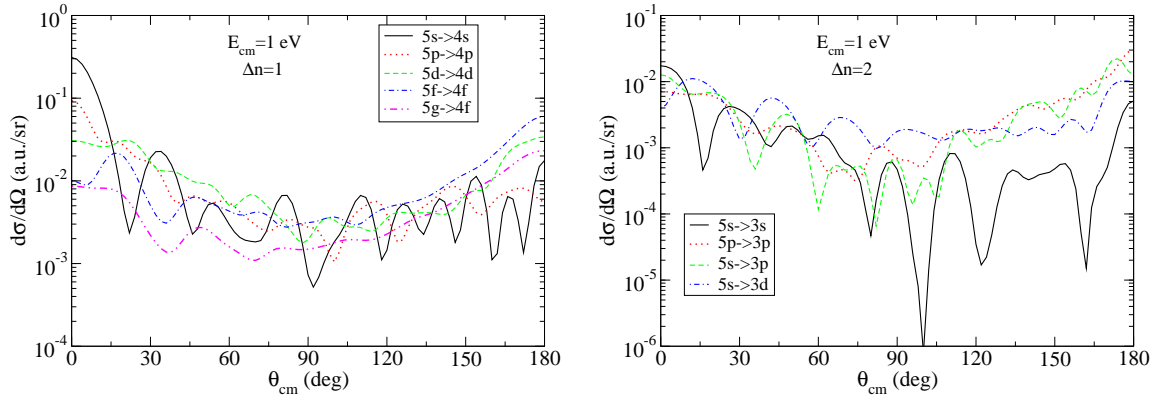

     \centerline{
     \includegraphics[width=0.45\textwidth,keepaspectratio]{fig3d.eps}\\
     \includegraphics[width=0.45\textwidth,keepaspectratio]{fig4d.eps}}
     \caption{Differential CD cross sections for the
  individual transitions with $\Delta n=1$ (left) and with $\Delta n=2$ (right)
  for $n=5$ at $E_{\rm {cm}}=1$~eV.}
     \end{figure}
We found that the angular distributions both of the individual and
$l$-averaged cross sections (excluding very low energies) are far
from isotropic: as a whole the scattering at $\theta_{cm} \lesssim
60^{\circ}$ and $\theta_{cm} > 120^{\circ}$ is noticeably
enhanced.
 The cross sections for  $ns\rightarrow n's$ transitions (see Fig.~7) have
 (as for elastic scattering) a more pronounced diffraction
 structure with sharp maxima and minima and a strong peak at zero angle
 as compared with the smoother angular dependence for other CD
 transitions.
 The increase of kinetic energy enhances asymmetry in
the  angular dependence of the $l$-averaged cross sections  and
decreases the role of the backward scattering (see Fig.~8).
     \begin{figure}[h]
    \parbox[c]{0.6\textwidth}{
     \includegraphics[width=0.6\textwidth,keepaspectratio]{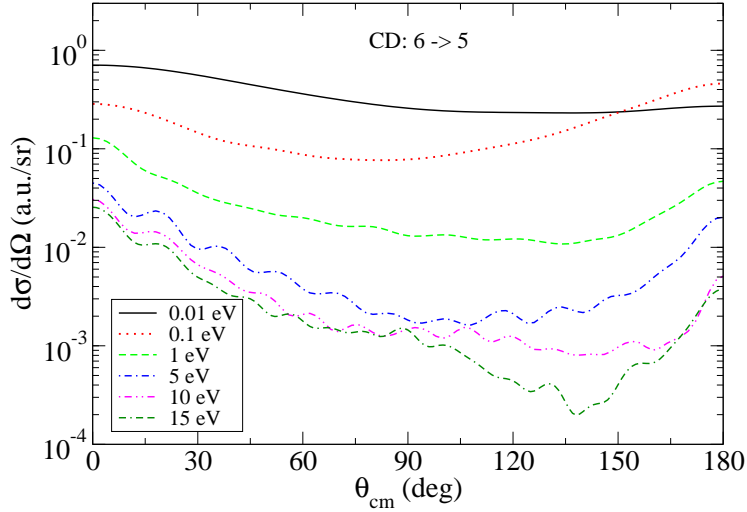}}
     \hfill
     \parbox[c]{0.35\textwidth}{\caption{The $l$-average differential cross sections
  for CD transition $6 \to 5$ at different energies.}}
    \end{figure}
\subsection{Hadronic atoms}
In order to illustrate the influence of the $ns$ state energy
shifts on the CD cross sections, we calculated the CD cross
sections for $(\pi p)_n + H$ collisions both with and without
taking energy shifts into account. The effect is the most
pronounced for the low-lying states and is illustrated in Fig.~9.
\begin{figure}[h!]
   \parbox[c]{0.5\textwidth}
   {\includegraphics[width=0.5\textwidth,keepaspectratio]{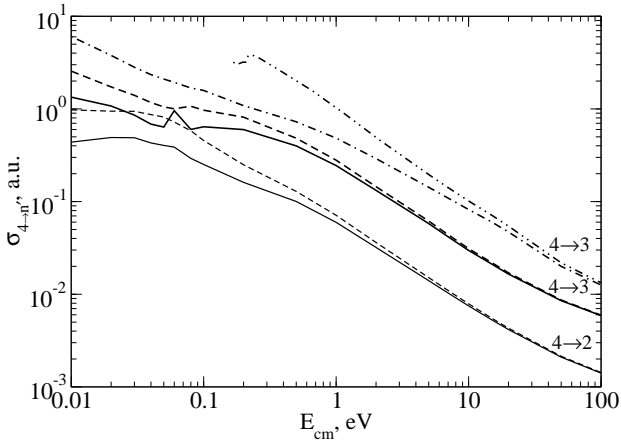}}
       \hfill
   \parbox[c]{0.45\textwidth}{ \caption{The $l$-averaged CD cross sections
 $\sigma^{l>0}_{4,3}$ (thick lines) and $\sigma^{l>0}_{4,2}$ (thin lines)
 for $(\pi p)_4 + H$ collisions calculated with (solid lines) and without
 (dashed lines) taking the $s$-states energy shifts into account.
 The CD cross sections $\sigma^{l=0}_{4\to 3}$ calculated both with (double-dot-dashed line)
 and without (dot-dashed line) taking $s$-state energy shifts into account are also shown.}}
     \end{figure}
 One can see that the maximal suppression due to the energy shift of
 $4s$ state is about two times
 at very low energy both for $4 \to 3$ and $4 \to 2$ transitions,
 while at $E_{cm} > 1$ eV does not exceed 15\% (for details see~\cite{12}).
  \begin{figure}[h!]
  \parbox[c]{0.5\textwidth}{\includegraphics[width=0.5\textwidth,keepaspectratio]{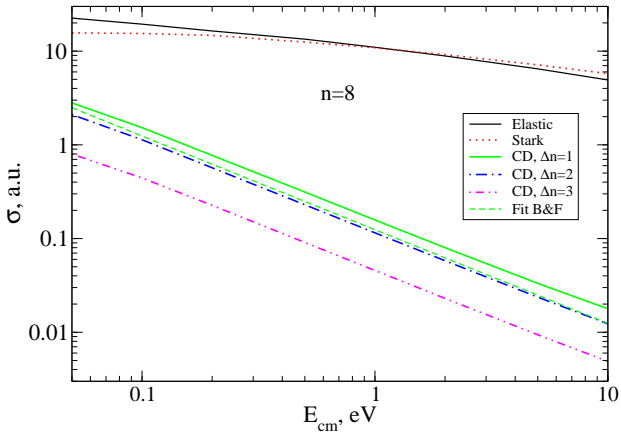}}
   \hfill
   \parbox[c]{0.45\textwidth}{\caption{The $l$-averaged CD cross sections  with
  $\Delta n=1,2,3$ for the collisions of the $p\bar{p}$ atom
   ($n=8$) with the hydrogen atom. The dashed line shows the fit
   used in cascade calculations for the transitions with
   $\Delta n =1$ and based on the mass-scaling of the results~\cite{4} for the
   muonic atom. The present results for the $l$-averaged elastic and
   Stark cross sections are shown for comparison.}}
     \end{figure}

The energy dependence of the CD cross sections for the collisions
of the $p\bar{p}$ atom
   ($n=8$) with the hydrogen atom obtained in
the CC approach is shown in Fig.~10 for $n=8$ and the different
values of $\Delta n$ =1, 2 and 3. The special features of these
cross sections are the following: the similar energy dependence
but sharper than that of the elastic scattering and Stark
transitions (see also in Fig. 10); the contribution of the
transitions with  $\Delta n > 1$ is comparable with the one for
$\Delta n$ =1 and is equal about 50\%. The effect of the $ns$
state shifts in the $l$-averaged CD cross sections is negligible
due to small statistical weight of the $ns$-state. In Fig.~10 we
also compare our results with those obtained in the semiclassical
model for the $\Delta n =1$ transition. The satisfactory agreement
is observed, but  this agreement is quite occasional and takes no
place for other $n$ values. The distribution over the final states
$n'$ is completely different from the SC results~\cite{4} as it
was illustrated in Fig.~10.

\section{Effect of closed channels}

The presented above results were obtained by the solution of the
close-coupling equations including all the open channels. Although
the channels with $n>n_0$ for $n_0\le 7$ and
 at energies less or $\sim 1$ eV  are strongly closed, they can
 essentially change the open-channel wave functions at short ranges
 determining CD process.
   \begin{figure}[h!]
  \parbox[c]{0.5\textwidth}{\includegraphics[width=0.5\textwidth,keepaspectratio]{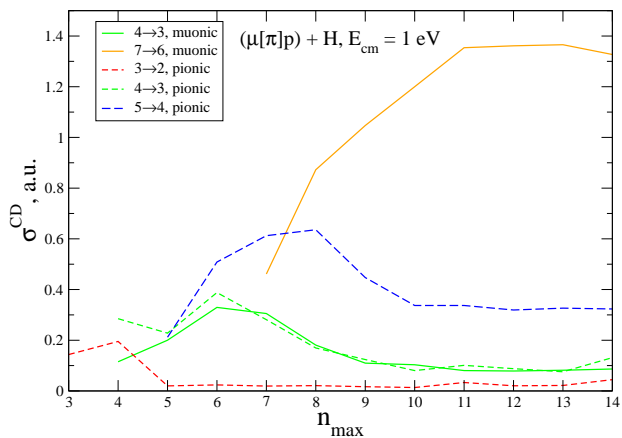}}
   \hfill
   \parbox[c]{0.45\textwidth}{\caption{Dependence of the CD cross sections
   $\sigma_{n_0\to n_0-1}$ for $(\mu^{-} p)$ and $(\pi^{-} p)$  atoms at $E_{\rm cm} =1$~eV on the
closed channels ($n_{max}$ is the maximal principal quantum number
of the included channels).}}
 \end{figure}
In the present studies we included in the calculations all the
open channels with $n<n_0$ and step by step added closed channels
with $n'>n_0,\,n'<n_{\rm max}$ to achieve the convergency of
results. In Fig.~11 some of our preliminary results are shown for
the $l$-averaged CD cross sections at $E_{\rm cm} =1$~eV.
According to our investigation, the closed channel effect on CD
cross sections depends crucially on the transition considered. As
it is seen from Fig.~11 in case of $(\mu^- p)$ atom the inclusion
in the basis set of the closed channels results in a more
pronounced effect for low-lying states. The investigations of
convergency with the increase of number of closed channels are
very time-consuming and are continuing now.
\section{Conclusion}
 The unified treatment of the elastic scattering, Stark
transitions and Coulomb deexcitation is presented within the
quantum-mechanical close-coupling approach. The differential and
integral cross sections for the above processes are calculated for
the excited muonic, pionic and antiprotonic hydrogen atoms with
$n=2-14$ and relative energies relevant to the cascade
calculations. The new results for CD process are obtained:
anisotropy of the angular distribution, substantial fraction of
$\Delta n > 1$ transitions up to $\sim 40\%$ (for $n\geq4$ at all
energies under consideration), and a proper threshold behaviour of
the CD cross-section. The calculated cross-sections are very
important for the kinetics of the atomic cascade and give a more
reliable theoretical input for the improved version of the cascade
model~\cite{13}.

 We are grateful to L.Ponomarev, L. Simons, G. Korenman,
 T.Jensen and V.Markushin for fruitful discussions.
This work was supported by Russian Foundation for Basic Research
(grant No. 06-02-17156).\\


\begin{thebibliography}{99}\itemsep -1mm
\bibitem {LB} M.Leon and H.A.Bethe, Phys.Rev. {\bf 127}, 636 (1962).
\bibitem {TH} T.P.Terrado and R.S. Hayano, Phys.Rev. C {\bf 55}, 73, (1977).

\bibitem {1}  V.P.Popov and V.N.Pomerantsev, Hyp. Interact. {\bf 101/102}, 133 (1996);
          {\bf 119}, 133 (1999); {\bf 119}, 137 (1999).
\bibitem {2} V.V.Gusev, V.P.Popov and V.N.Pomerantsev,
             Hyp. Interact. {\bf 119}, 141 (1999).
\bibitem {3} T.S.Jensen and V.E.Markushin, Eur.Journ.Phys. D, {\bf 19}, 165 (2001).
\bibitem {4} L.Bracci and G.Fiorentini, Nuovo Cim. {\bf 43A}, 9 (1978).
\bibitem {AAA} L.I.Ponomarev and E.A.Solov'ov, Yad.Fiz. {\bf 65}, 1615 (2002).
\bibitem {AAA1} A.V. Kravtsov and A.I. Mikhailov, Yad.Fiz. {\bf 69}, 395 (2006)
\bibitem {7} T.S.Jensen and V.E.Markushin, Eur.Journ.Phys. D, {\bf 21}, 261 (2001).
\bibitem {8} A.Badertscher {\em et al}., Europhys. Lett. {\bf 54} (3), 313 (2001).
\bibitem {9} G.Ya. Korenman, V.N. Pomerantsev, and V.P. Popov,
                 JETP Lett. {\bf 81}, 543 (2005); nucl-th/0501036.
\bibitem {10} V.P.Popov and V.N.Pomerantsev, nucl-th/051207.
\bibitem {11} V.N.Pomerantsev and V.P.Popov,  JETP Lett. {\bf 83}, 331 (2006).
\bibitem {12} V.N.Pomerantsev and V.P.Popov, Phys.~Rev. A {\bf 73}, 040501(R) (2006).
\bibitem {13} T.S.Jensen, V.P.Popov and V.N.Pomerantsev, arxiv:0712.3010 (2007).
\end{thebibliography}
\end{document}